\documentstyle[aps,prl]{revtex}

\newcommand{\Kr}[1]{\left( #1\right)}
\newcommand{\Ks}[1]{\left\langle #1\right\rangle}
\newcommand{\vf}{v_{\rm F}}
\newcommand{\kf}{k_{\rm F}}
\newcommand{\ve}{\varepsilon}
\newcommand{\lr}{\lambda_{\rm R}}

\begin{document}

\title{Rashba precession in quantum wires with interaction}

\author{Wolfgang H\"ausler}

\address{I.~Institut f\"ur Theoretische Physik der Universit\"at
Hamburg, Jungiusstr.~9, D-20355 Hamburg, Germany}

\date{\today}

\maketitle

\begin{abstract}
Rashba precession of spins moving along a one-dimensional quantum
channel is calculated, accounting for Coulomb interactions. The
Tomonaga--Luttinger model is formulated in the presence of
spin-orbit scattering and solved by Bosonization. Increasing
interaction strength at decreasing carrier density is found to
{\sl enhance} spin precession and the nominal Rashba parameter
due to the decreasing spin velocity compared with the Fermi
velocity. This result can elucidate the observed pronounced
changes of the spin splitting on applied gate voltages which
are estimated to influence the interface electric field in
heterostructures only little.
\end{abstract}
\vskip2pc
\noindent PACS~: 71.10.Pm, 71.70.Ej, 73.20.Dx


According to recent speculations spin could replace the electric
charge to carry information in future electronic devices
\cite{prinz}. The `spin transistor' proposed by Datta and Das
\cite{dattadas} might switch faster than traditional transistors
since during operation it avoids redistributing charges. The
idea is based on the Rashba effect \cite{rashba} which causes
spins to precess as they move along a heterostructure
\cite{bournel} so that the conductance depends on the final spin
orientation relative to the magnetization of the ferromagnetic
drain contact \cite{bauer,balents}. The strength of the Rashba
effect is proportional to the electric field acting perpendicular
to the electron plane which, when varied by a gate, changes the
final spin orientation and thus the transport properties of the
device. Advantage is taken of long spin coherence times and
lengths \cite{spincoherence}, found in semiconductors. In the
attempt of experimental realization considerable progress has
been achieved meanwhile to inject finite spin densities
\cite{spininjection}. Polarizations of 90\% have been reported
in GaAs \cite{molenkamp}.

Beating patterns varying with gate voltages have indeed been
observed in Shubnikov--de Haas (SdH) measurements
\cite{alphadec,grundler,matsuyama}. Estimates show, however,
that the the gate voltage adds an electric field contribution to
a much stronger intrinsic field at the interface which does not
suffice to explain the observed variations in spin splittings by
a factor of two. One should notice that the gate voltage not
only changes the strength of the electric field but at the same
time it also alters the carrier density in the heterostructure as
is directly monitored by the SdH oscillations. Without
interactions this would not change the Rashba precession within
the effective mass approximation \cite{narrowgap} but, since
particularly in semiconductors the strength of Coulomb
interactions change with density through the $r_{\rm
s}$--parameter, the question arises whether interactions
influence the Rashba precession. This is demonstrated in the
present Communications. In principle, the two effects of the
gate, changing field strength {\sl and} density, can be separated
experimentally by varying the voltage at a front gate and a back
gate independently \cite{grundler}.

Usually, Rashba spin precession is described as a band structure
effect, resulting from spin splitting \cite{dattadas,bournel},
as calculated in the original work by Rashba \cite{rashba} for
the homogeneous two-dimensional case. Such a single particle
approach, however, cannot account for electron-electron
interactions. Including many body effects requires going beyond
band structure theory. Particularly convenient
\cite{haldane,voit} and powerful \cite{yacoby} is the
Tomonaga--Luttinger (TL) model to incorporate interactions
exactly for the low energy and long wave length properties in
systems of one spatial dimension. From the practical point
of view, quantum wires seem most efficient to realize the spin
transistor since guiding the electron waves avoids angular
dispersion of momenta \cite{dattadas} and diminishes spin
scattering events \cite{bournel}. Spin injection into carbon
nanotubes, which are known to show pronounced TL behavior
\cite{balents} has been demonstrated \cite{alphenaar}. Here we
investigate a quantum wire with spin-orbit interaction in the
one-subband limit and apply the TL--model. We expect our main
result that interactions {\sl enhance} Rashba precession to
carry over to more channel situations. Indeed, a recent careful
analysis within improved versions of the RPA approximation has
revealed enhancement of the Rashba precession in the two
dimensional electron system \cite{raikh}. One might comprehend
these findings as a consequence of `repulsion' between the two
spin split bands by the interaction.

Spin splitting originates from spin--orbit coupling which in
narrow gap heterostructures, such as those based on InAs, is
dominated \cite{roessler} by the Rashba term
\begin{equation}\label{hso}
H^{\rm so}=\alpha(\sigma_xp_z-\sigma_zp_x)
\end{equation}
where $\sigma_{x,z}$ are Pauli matrices and $p$ is the electron
momentum in the $x$--$z$--plane of the heterostructure. Here,
we use coordinates indicated in the inset of Fig.~1. The Rashba
parameter $\alpha$ is mostly determined by the intrinsic
electric field perpendicular to the interface. Typical values
of 1---3$\times 10^{-11}$eVm \cite{alphanote} are reported for InAs
\cite{alphadec,grundler,matsuyama}.

Taking the confining potential in the lateral $z$--direction as
parabolic, the eigenenergies of
\begin{equation}\label{hrashba}
H=\frac{(p_z+m\alpha\sigma_x)^2+p_x^2}{2m}+\frac{m}{2}\omega_0^2z^2
-\alpha\sigma_zp_x
\end{equation}
determine the spin split single particle dispersion \cite{moroz}.
Here, $\omega_0$ is the subband energy and effective mass ($m$)
approximation is assumed. The eigenfunctions are plane waves of
momentum $k$ in $x$--direction along the wire and, to good
approximation, given as slightly modified oscillator functions
$\:\phi_n\:$ for the subband index $n$
\begin{equation}\label{psins}
\psi_{kns}(x,z)={\rm e}^{{\rm i}kx}\phi_n(z)
(\cos(m\alpha z)|s\rangle+{\rm i}\,\sin(m\alpha z)|\!-\!\!s\rangle)\;.
\end{equation}
On the wire axis the $\psi_{kns}(x,z)$ are spin polarized along
the $z$--direction ($s=\pm 1$) while the spins acquire a
non-zero out of plane $y$--component $\:\sim\sin(2m\alpha
z)=\sin(2\sqrt{\eta}z/\ell)\:$ away from the axis. This
describes a texture in the lateral spin density distribution.
Here, $\ell=(m\omega_0)^{-1/2}$ is the oscillator length and the
parameter $\:\eta=\frac{m}{\omega_0}\alpha^2=\frac{2\ve_{\rm
F}}{\omega_0}(\alpha/\vf)^2\:$ is introduced for later use
($\ve_{\rm F}$ is the Fermi energy above the subband edge and
$\vf$ the Fermi velocity). In the mostly used compounds
$\:\alpha/\vf<0.2\:$ \cite{grundler} (in GaAs it is even much
smaller), so that always $\:\eta<0.1\:$ in the lowest subband.

At $\:k=0\:$ different spin orientations are degenerate
but when $\:k\ne 0\:$ Rashba splitting occurs, yielding
\begin{equation}\label{ekns}
E_{k0s}=-m^*\alpha^{*2}+\frac{\omega_0}{2}+\frac{1}{2m^*}
(k+sm^*\alpha^*)^2\;,\;\;s=\pm 1
\end{equation}
for the energies of the lowest subband. Expression (\ref{ekns})
is obtained using the basis (\ref{psins}) after expanding
Laguerre polynomials arising from band mixing by the spin-orbit
term in powers of $\eta$. Up to the order ${\cal O}(\eta^{5/2})$
one can incorporate the effect of subband mixing, described by
$\eta$, into renormalized values for effective mass
$\:m^*=m(1+8\eta^2)\:$ and Rashba parameter
$\:\alpha^*=\alpha(1-\eta)\:$. Note that a similar reasoning
would be true also for other than parabolic dispersion relations
where effects of inter-subband mixing could still be incorporated
into renormalized values for the kinetic energy parameter and for
$\alpha$. The dispersion relation (\ref{ekns}) resembles the one
obtained in two dimensions \cite{rashba} but restricted to one
dimensional $k$--space, cf.\ Fig.~1. For higher subbands the
leading energy correction $\:\sim 2n\eta\:$, even in the strictly
parabolic dispersion case, depends on the subband index $n$ which
somewhat weakens the optimistic conjecture expressed by Datta and
Das \cite{dattadas} that the proposed spin transistor would not
loose sensitivity in multi-mode operation or at elevated
temperatures or voltages. The Rashba phases in higher subbands
differ slightly, $\sim\eta$, from the phase acquired in the
ground subband.

Without interactions the difference $\:2m^*\alpha^*\:$ in
momenta between $s=+1$ and $s=-1$ carriers, being independent of
Fermi energy, is the origin for the Rashba precession of spins
initially polarized e.g.\ along the $+x$--direction. After
traversing the Rashba distance $\lr=\pi/m^*\alpha^*$ the spin is
reversed. Contrary to the observations
\cite{alphadec,grundler,matsuyama}, Rashba splitting, and
therefore $\lr$, is independent of carrier density
$\:2\kf/\pi\:$ in (\ref{ekns}) at given interfacial electric
field \cite{narrowgap}; $\kf$ is the Fermi momentum. Also, we
note that the Fermi velocities of both spin components are
equal.

The term $\:\sim sm^*\alpha^*\:$ inside the brackets in
Eq.~(\ref{ekns}), shifting particle momentum, formally resembles
a vector potential associated with the magnetic flux through an
Aharanov--Bohm ring, leading there \cite{schmeltzer,loss} to a
persistent charge current in the ground state. Similarly, in
(\ref{ekns}) this term causes a non-zero persistent spin current
$\sim\Ks{J_{\sigma}}$ when periodic boundary conditions are
imposed \cite{meir}.

This similarity carries over to the TL--model for the quantum
wire at low energies which in Bose variables reads
\begin{equation}\label{hbose}
H=\sum_{\nu=\rho,\sigma}\frac{\pi}{4L}
\Kr{v_{\nu{\mbox{\tiny N}}}N_{\nu}^2+v_{\nu{\mbox{\tiny J}}}J_{\nu}^2}+
\Bigl(\sum_{q\ne 0}H_q\Bigr)-m^*\alpha^*\vf J_{\sigma}\;.
\end{equation}
Here, $\:N_{\nu}=N_{\nu\mbox{\tiny R}}+N_{\nu\mbox{\tiny L}}\:$
and $\:J_{\nu}=N_{\nu\mbox{\tiny R}}-N_{\nu\mbox{\tiny L}}\:$
denote particle numbers and currents, respectively, $\:N_{\rm
L/R}\:$ are the number of left/right going particles; $L$ is the
wire length. Topological excitations of $N_{\nu}$ or $J_{\nu}$
as well as density excitations at momentum $q$, described by
$H_q$, preserve charge ($\nu=\rho$) spin ($\nu=\sigma$)
separation (as already mentioned, here we have to high accuracy
equal velocities of both spins, cf.\ \cite{moroz}; this is not
the case, for example, in the presence of a Zeeman field
\cite{aoki}, when spin up and spin down velocities differ at the
Fermi energy due to the quadratic energy dispersion). The
interaction is exactly included in the TL--model by renormalized
values for its parameters such as the velocities
$v_{\nu{\mbox{\tiny N}}}$ and $v_{\nu{\mbox{\tiny J}}}$ which now
differ from $\vf$. In (\ref{hbose}) we have already omitted the
backscattering in spin sector originating from the exchange of
two electrons near two opposite Fermi points of opposite spins.
Such a term would couple the topological sector with the charge
and spin density excitations. For repulsive interactions back
scattering is known to be irrelevant at low energies
\cite{solyom}, also when persistent currents are present
\cite{loss}. Therefore, the $H_q$ will decouple, leaving the the
correlation exponents $K_{\nu}$ ($K_{\nu}\to 1$ without
interactions) and the power law decays of various correlation
functions, that can be calculated within the TL--model,
unaffected \cite{powernote} by the Rashba coupling $\sim\alpha$,
cf.\ Eq.~(\ref{psipsi}) below, contrary to the result
obtained in Ref.\ \onlinecite{moroz}. Spin-orbit coupling, and
therefore the Rashba term (the last term in (\ref{hbose})), does
not depend explicitly on the interaction.

Electrons injected at $x=0$ with spins polarized along the wire
axis will perform Rashba precession. Most directly this can be
monitored using the correlation function $\:f(x)=\frac{1}{2}
\Ks{(\Psi_{\uparrow}^{}(x)+\Psi_{\downarrow}^{}(x))
(\Psi_{\uparrow}^+(0)+\Psi_{\downarrow}^+(0))}\:$ where the
Fermi operator $\:\Psi_s(x)\propto {\rm e}^{-{\rm
i}\sqrt{\pi/2}(\phi_{\rho}(x)+s\phi_{\sigma}(x))} \sum_{r=\pm}
{\rm e}^{-{\rm i}r\sqrt{\pi/2}(\theta_{\rho}(x)+
s\theta_{\sigma}(x))}\:$ is expressed through the momentum
\[
\phi_{\nu}(x)=\sqrt{\frac{\pi}{2}}J_{\nu}\frac{x}{L}+
\sum_{q\ne 0}\phi_{\nu q}(x)
\]
and the density like Bose fields
\[
\theta_{\nu}(x)=\sqrt{\frac{\pi}{2}}N_{\nu}\frac{x}{L}+
\sum_{q\ne 0}\theta_{\nu q}(x)
\]
in the usual way \cite{haldane,voit}. The $\:q\ne 0\:$ components
yield power law decay while the $\:q=0\:$ components are
relevant for charge or spin stiffness and the persistent
current. The result for the desired correlation function is
\begin{eqnarray}\label{psipsi}
f(x)&=&-\frac{\kf}{\pi}|\kf x|^{-(K_{\rho}+1/K_{\rho}+K_{\sigma}+
1/K_{\sigma})/4}\\ \nonumber
&&\times\sin(|\kf x|)\;\cos(\frac{\pi}{2}j_{\sigma}x)\;.
\end{eqnarray}
The square $|f(L)|^2$ is proportional to the probability for spin
polarization in $+x$--direction at a distance $L$ from the source
and, by a similar reasoning as in \cite{dattadas}, to the current
accepted by an ideal ferromagnetic drain which is polarized
parallel to the source in $+x$--direction. A more sophisticated
full transport calculation has been carried out recently
\cite{balents} in the absence of spin-orbit coupling but would go
beyond the scope of the present work. The period in the slowly
oscillating last term in (\ref{psipsi}), $\:j_{\sigma}\equiv
\Ks{J_{\sigma}}/L=\frac{2}{\pi}m^*\alpha^*\vf/v_{\sigma{\mbox{\tiny
J}}}\:$, where the expectation value $\Ks{.}$ refers to the
ground state of (\ref{hbose}), determines the Rashba length
\begin{equation}\label{lrint}
\lr=\frac{\pi}{m^*\alpha^*}\frac{v_{\sigma{\mbox{\tiny J}}}}{\vf}\;.
\end{equation}
This quantity depends on the velocity of the spin current
$v_{\sigma{\mbox{\tiny J}}}$ and is now altered compared to its
value in the absence of interactions. We note, that in charge
sector the corresponding velocity $v_{\rho{\mbox{\tiny J}}}=\vf$
remains constant as a consequence of Galilei invariance of the
quantum wire \cite{voit} at carrier densities much smaller than
the inverse lattice constant of the underlying semiconductor
lattice. Galilei invariance does not hold in spin sector where
particles of opposite spins moving in opposite directions will
experience some drag by the interaction. Therefore,
$v_{\sigma{\mbox{\tiny J}}}$ differs from $\vf$.

Rigorous relations hold \cite{haldane} among the velocities
$\:v_{\nu{\mbox{\tiny J}}}=v_{\nu}K_{\nu}=v_{\nu{\mbox{\tiny
N}}}K_{\nu}^2\:$ in TL--liquids, determined by the correlation
exponents. Here, $v_{\nu}$ is the velocity of density
excitations at small $q$. In many cases spin rotation
invariance holds and fixes $K_{\sigma}=1$. Rashba coupling
breaks this SU(2) invariance. Fortunately, the corresponding
energy $\:2m^*\alpha^*\vf\:$ is small compared to the Fermi
energy (or to the typical Coulomb energy $v_{\rho{\mbox{\tiny
N}}}/\vf$) so that $K_{\sigma}$ will deviate from unity at most
slightly and $\:v_{\sigma{\mbox{\tiny J}}}\approx
v_{\sigma}\approx v_{\sigma{\mbox{\tiny N}}}\:$ to good
accuracy. Much more important than the differences between
those three velocities are their deviation from $\vf$ occurring
already in the SU(2) symmetric case as seen for example in the
Hubbard chain \cite{schulz90}. How long range interactions
alter $v_{\sigma}/\vf$ has been investigated perturbatively
\cite{schulz93} and, recently, by extensive quantum Monte Carlo
studies \cite{creffield}. Starting from values close to unity
at high carrier densities, $\:r_{\rm s}<0.5\:$, $v_{\sigma}/\vf$
can drop below 0.5 when $\:r_{\rm s}\gtrsim 1.4\:$. Those
densities are easily reached in present day quantum wires
\cite{yacoby}. In the limit of very small particle densities
$2\kf/\pi$ the spin velocity is expected to vanish like
$\:v_{\sigma}\sim \kf^2\:$.

Thus, according to (\ref{lrint}), Rashba precession is {\sl
enhanced} by repulsive interactions. Qualitatively, one might
understand this result on the mean field level as a consequence
of the Fock contribution precipitating repulsions between
opposite rather than same spins. This enhances spin splitting
and the nominal value of the Rashba parameter. In principle
this argument applies also to situations with more than one
subband occupied or even two dimensional heterostructures.
Also in those cases we would expect that Coulomb interactions
amplify Rashba precession as it has been confirmed in Ref.\
\onlinecite{raikh}.

The amount of Rashba enhancement expected at given carrier
density can be measured independently through the Zeeman spin
susceptibility $\:\chi=-(\partial^2E_0/\partial
B^2)/L=(L\:\partial^2E_0/\partial N_{\sigma}^2)^{-1}=2/\pi
v_{\sigma{\mbox{\tiny N}}}\:$ by monitoring the `exchange
enhancement' of the effective $g$--factor; $E_0$ is the ground
state energy of (\ref{hbose}). In InAs the magnetic energy $B$
translates quite accurately as 1 Ry per Tesla when taking the
$g$--factor \cite{ensslin} as $|g|=13$. Assuming again
$K_{\sigma}\approx 1$ yields
\begin{equation}
\lr=2/m^*\alpha^*\vf\chi\;.
\end{equation}

In conclusion, we have established a theory beyond describing
Rashba precession as a single particle band structure effect. 
We have considered a quantum wire and the TL--model to
incorporate interactions exactly. Increasing repulsion between
carriers along with decreasing particle densities is found to
reduce the Rashba length $\lr$ over which spins complete cycles
as they move along the wire. Accordingly, the nominal value of
the Rashba parameter increases, as determined by Shubnikov--de
Haas measurements. This is demonstrated to be a consequence of
decreasing spin velocities. The latter could be measured
independently through the magnetic susceptibility w.r.t.\ a
Zeeman field. Contrary to the relatively small influence of
gates on the strength of the interfacial electric field this
interaction induced contribution can explain variations of $\lr$
by a factor of 2. It would be valuable to experimentally
separate the influence of the field strength from the carrier
density by applying voltages independently to a front and a back
gate \cite{grundler}. Moreover, the front gate could screen the
long range part of the Coulomb interaction and thereby serve to
vary the microscopic interaction strength.

I thank Charles Creffield, Dirk Grundler, Allan MacDonald,
Ulrich Merkt, and, especially, Ulrich Z\"ulicke for valuable
discussions, and M.\ E.\ Raikh for drawing my attention to
the work of Ref.~\onlinecite{raikh}.

\begin{raggedright}\end{raggedright}
\newpage\vspace*{22cm}\hspace*{-3cm}\includegraphics{figure1}

\vspace*{-7cm}
\hspace*{5cm}FIGURE 1 

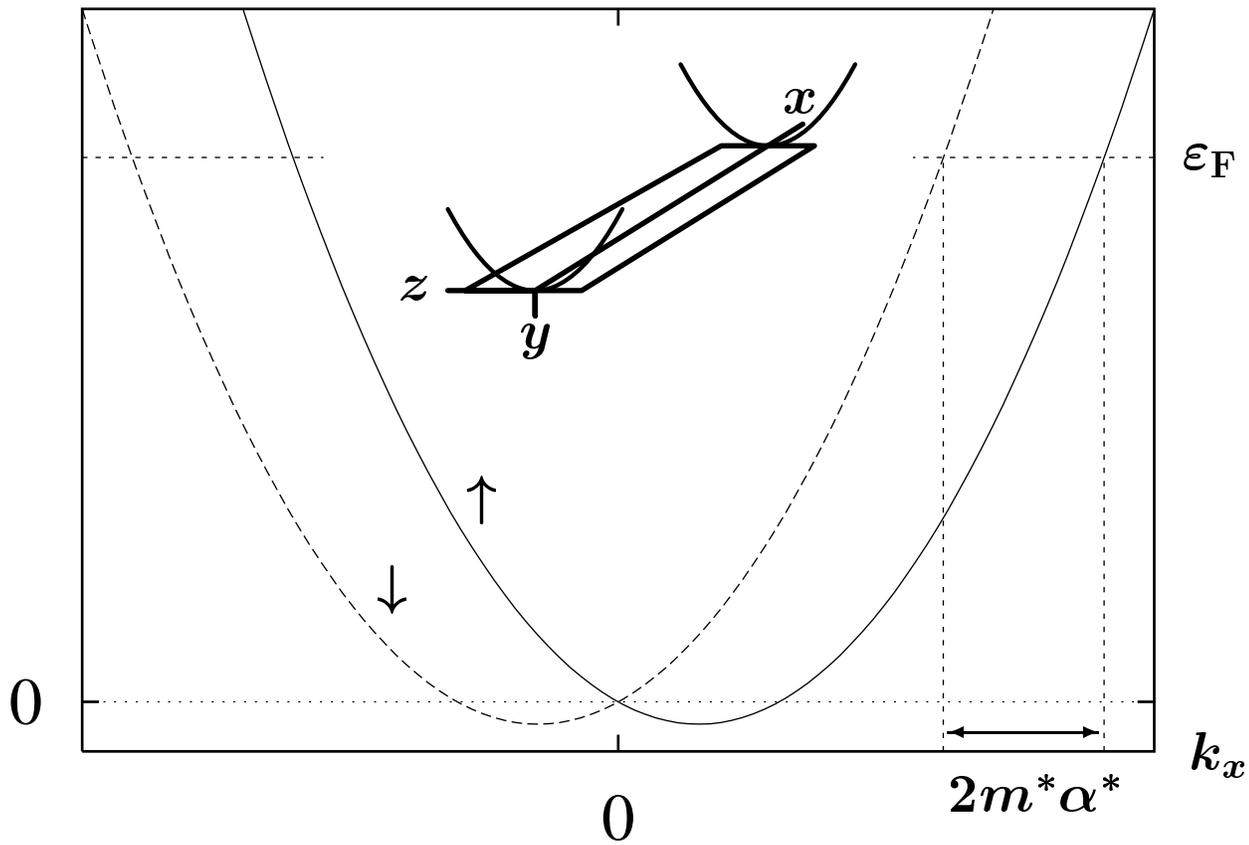
Fig.~1: Energy dispersion in the lowest spin split subband of a
quantum wire with parabolic confining potential. On the wire axis
the spins $\:s=\uparrow,\downarrow\:$ are polarized in the
plane of the heterostructure.


\begin{thebibliography}{10}
\bibitem{narrowgap} In narrow gap semiconductors with almost
linear single particle dispersion the Rashba precession would
decrease with decreasing electron density which is opposite to
what is predicted here as a consequence of interaction effects.
\bibitem{alphanote} In the present work we consider $\alpha$ as being
defined through (\ref{hso}) while $2\lr$ is the {\sl observed}
spatial length carriers have to traverse to perform one cycle in
spin. Thus $\lr$ may depend on particle density while $\alpha$
does not.
\bibitem{powernote} Actually, there is a non-zero change of the
TL--parameters also in the charge sector, describing density
excitations, due to the Rashba term since the Coulomb matrix
elements are slightly changing with the wave functions
(\ref{psins}). This effect, however, is much smaller than the
influence of a changing carrier density and can safely be
ignored.
\end{thebibliography}
\end{document}